\documentclass[12pt,showpacs]{revtex4}

\usepackage{graphicx}
\begin{document}
\title{Stationary cantilever vibrations in the oscillating
cantilever-driven adiabatic reversals -- magnetic resonance
force microscopy technique}
\author{G.P. Berman$^1$, D.I. Kamenev$^1$, 
and V.I. Tsifrinovich$^2$}
\affiliation{$^1$Theoretical Division and Center for Nonlinear Studies,
Los Alamos National Laboratory, Los Alamos, New Mexico 87545}
\affiliation{$^2$IDS Department, Polytechnic University,
Six Metrotech Center, Brooklyn, New York 11201}
\vspace{3mm}
\begin{abstract}
We consider theoretically the novel technique in magnetic resonance force
microscopy which is called ``oscillating cantilever-driven
adiabatic reversals''. We present analytical and numerical
analysis for the stationary cantilever vibrations in this technique.
For reasonable values of parameters we estimate the resonant
frequency shift as 6Hz per the Bohr magneton. We analyze also the regime
of small oscillations of the paramagnetic moment near the transversal
plane and the frequency shift of the damped cantilever vibrations.
\end{abstract}
\pacs{03.67.Lx,~03.67.-a,~76.60.-k}
\maketitle
\section{Introduction}
Magnetic resonance force microscopy (MRFM) based on a cyclic adiabatic
inversion (CAI) is considered as one of the most promising
roads to the ultimate goal of a single-spin detection in solids
(see, for example,~\cite{2,3}). Typically CAI is generated using the
frequency modulation of the external radio-frequency ({\it rf}) field.
In this case, a paramagnetic moment of a sample follows the effective
magnetic field in the rotating system of coordinates (RSC), 
and influence the cantilever vibrations.

Recently a new technique called ``oscillating cantilever-driven
adiabatic reversals'' (OSCAR) has been suggested and implemented
in~\cite{1}. In this technique, the cantilever driven by an external
force causes the CAI of the paramagnetic moment of a sample.
The back reaction of the paramagnetic moment causes the frequency
shift of the cantilever vibrations, which is supposed to be detected.
The main purpose of this paper is a theoretical analysis of the stationary
vibrations of the cantilever in the OSCAR technique. Our consideration
is based on the classical equations of motion for the spin-cantilever
system.

The paper is organized as follows. In Sec. II we introduce the model.
The linear OSCAR regime is considered in Sec. III, and nonlinear regime
is analyzed in Sec. IV. A perturbative approach and numerical results
are presented in Sec. V. In Sec. VI, we analyze the damped oscillations of
the cantilever in the absence of the external force. In Sec. VII we give a
brief summary of our results.

\begin{figure}
\centerline{\includegraphics[width=13cm,height=7cm]{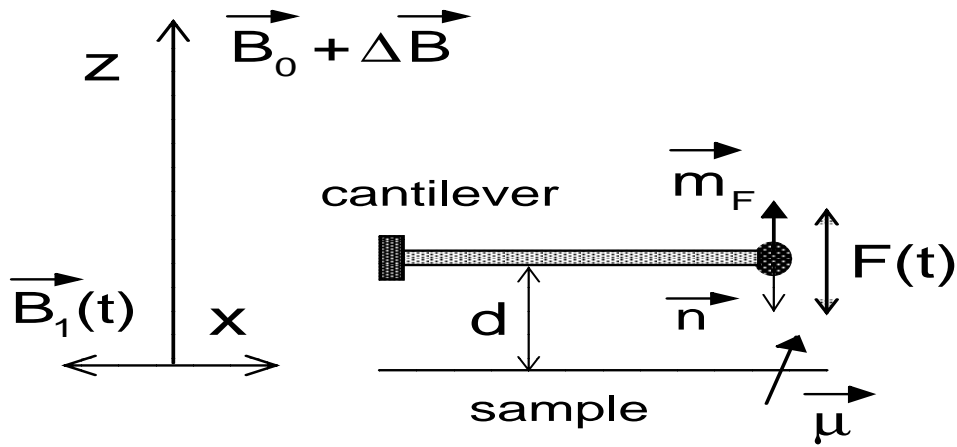}}
\vspace{-6mm}
\caption{A schematic setup of the system under consideration.
$\vec{B}_0+\Delta\vec{B}$ is the uniform permanent magnetic field,
$\vec{B}_1$ is the rotating {\it rf} magnetic field, $F(t)$ is an
external force acting on the cantilever in the $z$-direction,
$\vec{m}_F$ is the magnetic moment of the ferromagnetic particle,
$\vec{\mu}$ is the magnetic moment of the paramagnetic cluster,
$d$ is the equilibrium distance between the center of the ferromagnetic
particle and the cluster.}
\label{fig:1}
\end{figure}

\section{Hamiltonian and equations of motion}
A schematic setup of the studied system is shown in Fig.~1.
A spherical ferromagnetic particle with magnetic moment, $\vec{m}_F$,
is attached to the cantilever tip. A small paramagnetic cluster with
magnetic moment, $\vec{\mu}$, which must be detected, is placed on the
surface of non-magnetic sample beneath the tip of the cantilever.
The whole system is placed into the high permanent magnetic field,
$\vec{B}_0+\Delta\vec{B}$, oriented in the positive $z$-direction.
The external force, $F(t)$, drives the cantilever vibrations
along the $z$-axis. The transversal rotating magnetic field,
$\vec{B}_1(t)$, is applied to the paramagnetic cluster. We place the
origin of our coordinate system at the equilibrium position of the
cantilever tip.

We consider the cantilever tip as an oscillator with the effective mass,
$m^*$, and the effective spring constant, $k_s$. The classical Hamiltonian
for the cantilever with the ferromagnetic particle and the paramagnetic
cluster has the form,
\begin{equation}
\label{1}
{\cal H}={{p_z^2}\over{2m^*}}+{{k_sz^2}\over{2}}-zF(t)-{{\mu_0m_F}
\over{2\pi(d+z)^3}}\mu_z-
\vec{\mu}(\vec{B}_0+\Delta\vec{B}+\vec{B}_1),
\end{equation}
where $p_z$ and $z$ are the momentum and coordinate of the cantilever tip,
$\mu_0$ is the permeability of the free space. Putting
$F(t)=F_0\cos(\nu t)$ and taking into consideration the finite quality
factor, $Q$, of the cantilever we write the equation of motion for the
cantilever,
\begin{equation}
\label{2}
\ddot z+\omega^2_cz+{{q\mu_z}\over{(d+z)^4}}+{{\omega_c}\over{Q}}\dot z=
f_0\cos(\nu t),
\end{equation}
where $\omega_c=(k_s/m^*)^{1/2}$ is the unperturbed cantilever frequency,
$f_0=F_0/m^*$, and,
\begin{equation}
\label{3}
q={{3\mu_0m_F}\over{2\pi m^*}}.
\end{equation}

Next, we assume that the {\it rf} field $\vec{B}_1$ rotates in the
$(x\,,y)$ plane with the frequency,
\begin{equation}
\label{4}
\omega_0=\gamma[B_0+B_d(0)].
\end{equation}
Here $\gamma$ is the gyromagnetic ratio of the paramagnetic cluster,
$B_d(z)$ is the dipole magnetic field produced by the ferromagnetic
particle at the point of location of the paramagnetic cluster,
\begin{equation}
\label{5}
B_d(z)={\mu_0m_F\over 2\pi(d+z)^3},
\end{equation}
and $B_d(0)$ is the value of $B_d(z)$ at the equilibrium position
of the cantilever, $z=0$.
The equation of motion for the paramagnetic moment $\vec \mu$ in the RSC
has the form,
\begin{equation}
\label{6}
\dot{\vec{\mu}}=\gamma[\vec{\mu}\times \vec{B}_{eff}].
\end{equation}
Here $\vec{B}_{eff}$ is the effective magnetic field in the RSC
with the $x$-component $B_1$ and the $z$-component
$\Delta B+B_d'(z)$, where $B_d'(z)$ is the oscillatory part
of the dipole field produced by the ferromagnetic particle on the
cluster:
\begin{equation}
\label{7}
B_d'(z)=B_d(z)-B_d(0).
\end{equation}

\section{The linear OSCAR regime: small oscillations of $\vec\mu$}
\label{sec:qualitative}
In this section we consider the linear OSCAR regime. Suppose that initially
an auxiliary $\pi/2$-pulse changes the direction of the
paramagnetic moment, $\vec\mu$, from $+z$ to $+x$ of the RSC. We also assume that the
oscillatory part of the dipole field, $B_d'(z)$, is small compared
to the {\it rf} field, $B_1$. Certainly we assume that the unperturbed
cantilever frequency, $\omega_c\ll \gamma B_1$, to keep the conditions of
CAI. In quasi-static approximation a paramagnetic moment, $\vec \mu$,
follows the effective field $B_{eff}$. Putting in (\ref{6})
$\dot{\vec\mu}=0$, we obtain for $|z|\ll d$:
\begin{equation}
\label{8}
\mu_x(t)\approx \mu,~\mu_y(t)=0,~\mu_z(t)=
{\mu\over B_1}\left[\Delta B-{3\mu_0m_Fz(t)\over 2\pi d^4}\right].
\end{equation}
These equations describe small (linear) oscillations of $\vec\mu$ near
the $x$-axis. Substituting the last expression in Eq.~(\ref{8}) into
Eq.~(\ref{2}) we derive an approximate equation for the cantilever oscillations,
\begin{equation}
\label{9}
\ddot z+\omega^{*2}_cz+{\omega^*_c\over Q^*}\dot z=
f_0\cos(\nu t).
\end{equation}
Here,
\begin{equation}
\label{10}
\omega^*_c=\omega_c+\Delta\omega_c,~\Delta\omega_c=
-{3\mu_0m_F\mu\over \pi m^*\omega_cB_1d^5}
\left(\Delta B+{3\mu_0m_F\over 8\pi d^3}\right),~Q^*=
Q\left(1+{\Delta\omega_c\over\omega_c}\right).
\end{equation}

Equation (\ref{9}) describes the motion of the linear oscillator
with the effective frequency, $\omega^*_c$, and the effective quality factor,
$Q^*$. Due to the back reaction of the paramagnetic moment on the
cantilever the effective frequency and the quality factor of the cantilever
depend on the permanent magnetic field, $\Delta B$,
(in our approximation, $\Delta B\ll B_1$).
If $\Delta B>-3\mu_0m_F/(8\pi d^3)$, then both the frequency and
the quality factor of the cantilever decrease. In the opposite case they
increase.

\section{Nonlinear adiabatic regime: adiabatic reversals of $\vec\mu$}
To increase the back reaction of $\vec\mu$ it is important to provide
large oscillations (adiabatic reversals) of the paramagnetic moment.
In this section we consider stationary vibrations of the cantilever
in the nonlinear OSCAR regime. It is convenient to write the
equations of motion in the dimensionless form:
$$
Z''+Z+{{\lambda M_z}\over{(1+\alpha Z)^4}} +
{{1}\over{Q}}Z'={{1}\over{Q}}\cos[(1+\rho)\tau+\vartheta_0],
$$
\begin{equation}
\label{11}
{M_x}'=(\delta-\chi Z)M_y,
\end{equation}
$$
{M_y}'=\varepsilon M_z-(\delta-\chi Z) M_x,
$$
$$
{M_z}'=-\varepsilon M_y,
$$
where we introduced the dimensionless time $\tau=\omega_ct$;
prime means a differentiation over $\tau$, $Z=z/A$ is the dimensionless
coordinate, $A=f_0Q/\omega_c^2$ is the unperturbed (in the absence of the magnetic moment ${\vec M}$) amplitude of the
stationary cantilever vibrations in the resonant regime
(when $\omega=\omega_c$), $\vec M=\vec\mu/\mu$ is the dimensionless
magnetic moment, $\delta=\gamma\Delta B/\omega_c$.
The parameter $\alpha=A/d$ is small, $\alpha\sim 0.01$.
The dynamics is controlled by the following dimensionless parameters:
$$
\lambda={{3\mu_0m_F\mu}\over{2\pi d^4QF_0}},
$$
\begin{equation}
\label{12}
\chi={{3\gamma\mu_0m_FQf_0}\over{2\pi \omega_c^3d^4}},
\end{equation}
$$
\varepsilon={{\gamma B_1}\over{\omega_c}},\qquad \rho=\nu/\omega_c-1.
$$

Suppose that the paramagnetic moment, $\vec M$, points initially in the direction of
the effective magnetic field, $\vec B_{eff}$, and the cantilever
points in the opposite direction, $Z(0)=-1$.
In this case the quasi-static motion of $\vec M$ is given by the expressions,
$$
{M_x(\tau)}={\varepsilon\over\sqrt{\varepsilon^2+(\delta-\chi Z)^2}} ,
$$
\begin{equation}
\label{13}
{M_y}=0,
\end{equation}
$$
{M_z(\tau)}={\delta-\chi Z\over\sqrt{\varepsilon^2+(\delta-\chi Z)^2}}.
$$
Substituting (\ref{13}) into the first equation in (\ref{11}) we
obtain the nonlinear equation for $Z$:
\begin{equation}
\label{14}
Z''+Z-{\lambda\chi Z\over\sqrt{\varepsilon^2+(\chi Z)^2}}
+{1\over{Q}}Z'={1\over Q}\cos[(1+\rho)\tau+\vartheta_0],
\end{equation}
where we neglected the term $\alpha Z$ in the denominator in the
third term in the left-hand side and put $\delta=0$. The third term in  Eq.~(\ref{14}) corresponds to the modification of the potential energy of the cantilever, due to the interaction with the magnetic moment, $\vec M$, by the value,
$$
\delta U(Z)=-{{\lambda}\over{2\chi}}\sqrt{\varepsilon^2+(\chi Z)^2}.
$$

Now, we present an approximate ``semi-quantitative'' analysis of the
stationary oscillations described by Eq.~(\ref{14}).
The solution for the stationary driven oscillations of the cantilever,
described by Eq.~(\ref{14}), can be written in the form,
\begin{equation}
\label{15}
Z=a(\rho)\sin[(1+\rho)\tau+\vartheta_0].
\end{equation}
We define the frequency shift which corresponds to the shift
of the maximum, $a_{max}(\rho_1)$, of the
amplitude, $a=a(\rho)$, caused by the ferromagnetic sample.
In order to estimate $a_{max}$ we replace,
$$
Z^2\sim\sin^2[(1+\rho)\tau+\vartheta_0]=
\frac 12\{1-\cos[2(1+\rho)\tau+2\vartheta_0]\}\rightarrow\frac 12,
$$
in the denominator in the third term in Eq.~(\ref{14}), and neglect the term
$\cos[2(1+\rho)\tau+2\vartheta_0]$ because it is non-resonant.
Then Eq.~(\ref{14}) takes the form,
\begin{equation}
\label{one_eq1}
Z''+\left[1-{\lambda\chi\over\sqrt{\varepsilon^2+\frac{\chi^2}2}}\right]Z
+{{1}\over{Q}}Z'={{1}\over{Q}}\cos[(1+\rho)\tau+\vartheta_0].
\end{equation}
The position, ${\rho}_1$, of the maximum of the amplitude,
$a_{max}$, of the driven oscillations (the frequency shift) is,
\begin{equation}
\label{16}
{\rho}_1-\rho_0=
-{\lambda\chi\over 2\sqrt{\varepsilon^2+\frac{\chi^2}2}}
\approx - {\lambda\over\sqrt 2},
\end{equation}
where $\rho_0=-1/4Q^2$ is the position of the maximum of the amplitude
in the absence of the paramagnetic sample, and we suppose that $\chi\gg\varepsilon$.

For estimation of the value of the frequency shift the following
parameters where used: $D=1.5\times 10^{-7}$m is the diameter of the
ferromagnetic particle with the volume $V=1.8\times 10^{-21}$ m$^3$,
$\mu_0m_F/V\approx 1.1$T, $k_c\approx 10^{-3}$N/m,
$\omega_c/2\pi\approx 10^5$Hz, $A\approx 1$nm, $d\approx 100$nm,
$B_1\approx 10^{-3}$T. For these values of parameters we obtain,
\begin{equation}
\label{17}
\varepsilon\approx 280,~\chi\approx 2.5\times 10^{3},~\lambda\approx
8.5\times10^{-5} (\mu/\mu_B),~\alpha=0.01,
\end{equation}
where $\mu/\mu_B$ is the paramagnetic moment expressed in units
of the Bohr magneton. The corresponding frequency shift is,
\begin{equation}
\label{18}
{\rho}_1-\rho_0\approx -6\times 10^{-5} (\mu/\mu_B).
\end{equation}
This gives the frequency shift -6 Hz per one Bohr magneton.

\section{Perturbation approach}
The qualitative estimation presented above can be supported
by application of the approach based on the perturbation theory developed by Bogoliubov and Mitropolskii in~\cite{4}. We look for the solution
of Eq.~(\ref{14}) in the form,
\begin{equation}
\label{19}
Z=a(\tau)\cos[\psi]+\lambda u_1(a,\,\psi),
\end{equation}
where $\psi=(1+\rho)\tau+\vartheta(\tau)$. The function
$u_1(a,\,\psi)$ is the sum of the Fourier terms with the
phases $3\psi$, $5\psi$, $7\psi$, $\dots$.
The amplitudes of these terms decrease with increasing the Fourier
number, $n$, as $1/(2n+1)^2$. The first non-vanishing
term is small and equals to
$u_1(a,\,\psi)\approx 0.02\cos(3\psi)$.
This allows us to  neglect the contribution of $u_1(a,\,\psi)$ into
the expression for $Z$ in Eq.~(\ref{19}).

The slow varying amplitude, $a(\tau)$, and the phase,
$\vartheta(\tau)$, in the first order of the perturbation theory
satisfy the two coupled differential equations,
\begin{equation}
\label{20}
{da\over d\tau}=-\frac\lambda{2\pi}\int_0^{2\pi}
{\chi a\cos\psi\sin\psi d\psi\over
\sqrt{\varepsilon^2+(\chi a\cos\psi)^2}}-{a\over 2Q}-
{1\over Q(2+\rho)}\sin\vartheta,
\end{equation}
\begin{equation}
\label{21}
{d\vartheta\over d\tau}=-{1\over 8Q^2}-\rho-
\frac\lambda{2\pi a}\int_0^{2\pi}
{\chi a\cos^2\psi d\psi\over
\sqrt{\varepsilon^2+(\chi a\cos\psi)^2}}-
{1\over aQ(2+\rho)}\cos\vartheta.
\end{equation}
Note, that the integral in the right-hand side of Eq.~(\ref{20})
is equal to zero.
The integral in the right-hand side of Eq.~(\ref{21}) can be
expressed through the elliptic integrals as (see Appendix),
\begin{equation}
\label{22}
4\int_0^{\pi/2}
{\chi a\cos^2\psi d\psi\over
\sqrt{\varepsilon^2+(\chi a\cos\psi)^2}}=
4\left[\frac 1kE(k)-p^2kK(k)\right],
\end{equation}
where $k=1/\sqrt{1+p^2}$, $K(k)$ and $E(k)$ are the complete
elliptic integrals, respectively, of the first and second kind,
$p=\varepsilon/(a\chi)$.
When $p^2\ll 1$ one can decompose $K(k)$ and $E(k)$ as,
\begin{equation}
\label{23}
K(k)\approx C+(C-1){{k'}^2\over 4}+\dots,\qquad
E(k)\approx 1+ \left(C-\frac 12\right){{k'}^2\over 2}+\dots,
\end{equation}
where ${k'}^2=1-k^2\approx p$, $C=\ln(4/k')\approx\ln(4/p)$.
From Eqs.~(\ref{22}) and (\ref{23}) we find the value of the integral
in Eq.~(\ref{21}) for $p\ll 1$,
\begin{equation}
\label{24}
-\frac\lambda{2\pi a}\int_0^{2\pi}
{\chi a\cos^2\psi d\psi\over
\sqrt{\varepsilon^2+(\chi a\cos\psi)^2}}\approx
-\frac{2\lambda}{\pi a}\left[1-{p^2\over 4}
\left(2\ln\frac 4p-1\right)\right].
\end{equation}

Substituting Eq.~(\ref{24}) to Eq.~(\ref{21}) we obtain,
\begin{equation}
\label{25}
{da\over d\tau}=-{a\over 2Q}-{1\over Q(2+\rho)}\sin\vartheta,
\end{equation}
$$
{d\vartheta\over d\tau}=-{1\over 8Q^2}-\rho-
\frac{2\lambda}{\pi a}\left[1-{p^2\over 4}
\left(2\ln\frac 4p-1\right)\right]
-{1\over aQ(2+\rho)}\cos\vartheta.
$$

We now calculate
the position of the maximum of the amplitude, $a(\rho)$, in the
stationary regime of driven oscillations using Eq.~(\ref{25}),
and compare it with the results obtained in Sec.~\ref{sec:qualitative}.
In the regime of driven oscillations $a=const$, $\vartheta=const$,
and we must solve
the system of two equations (\ref{25}), where
$da/d\tau=0$ and $d\vartheta/d\tau=0$. Canceling the phase, $\vartheta$,
we have,
\begin{equation}
\label{26}
{1\over a^2(2+\rho)^2}=\frac 14+Q^2\left(
{1\over 8Q^2}+\rho+{2\lambda\over\pi a}\right)^2,
\end{equation}
where we neglected the term proportional to $p^2\ll 1$.
The amplitude, $a$, can be written as $a=1+\beta$, where $\beta\ll 1$,
so that,
\begin{equation}
\label{27}
{1\over a(2+\rho)}={1\over(1+\beta)(2+\rho)}\approx
\frac 12\left(1-\beta-\frac\rho 2\right).
\end{equation}
Taking the square root from the both sides of Eq.~(\ref{26}) and
using Eq.~(\ref{27}) we obtain,
\begin{equation}
\label{28}
-\beta-\frac\rho 2\approx 2Q^2\left(
{1\over 8Q^2}+\rho+{2\lambda\over\pi}\right)^2,
\end{equation}
where we put $a\approx 1$ in the denominator of the term proportional
to $\lambda$ (i.e. we neglected the term of the order of $\beta\lambda$).
The maximum of the function, $\beta=\beta(\rho)$, can be found from the
condition $d\beta(\rho_1)/d\rho=0$ which yields,
\begin{equation}
\label{29}
\rho_1=-{1\over 4Q^2}-{2\lambda\over\pi}.
\end{equation}
This is approximately the same value as that given by Eq.~(\ref{17}),
obtained from the qualitative considerations.
The second term in Eq.~(\ref{29}) describes the influence of the
paramagnetic moment reversals on the resonance frequency of the
cantilever.

\vspace{-1mm}
\begin{figure}
\setcounter{figure}{1}
\centerline{\includegraphics[width=13cm,height=7cm]{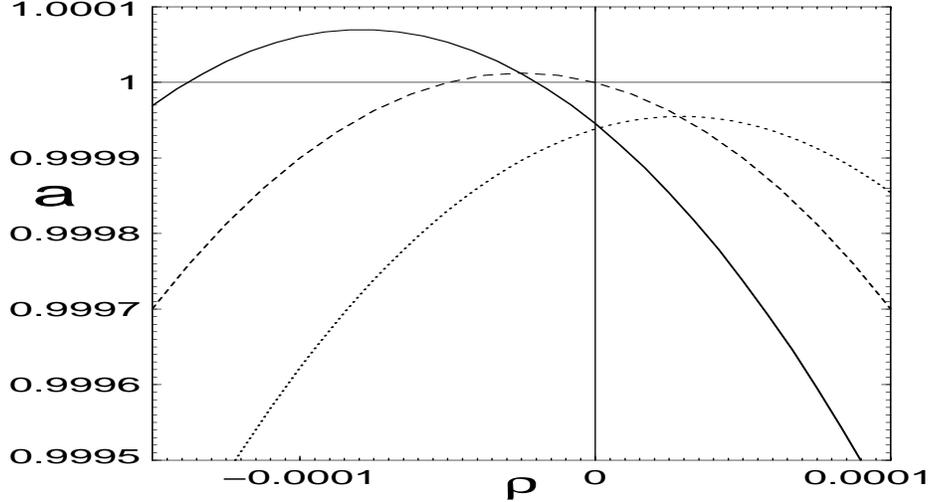}}
\caption{The dependence of the amplitude of the driven oscillations
of the cantilever on the frequency detuning, $\rho$, obtained
using numerical solution of exact equations of motion (\ref{11}).
Solid line corresponds to the initial conditions (\ref{30}) and
the values of the parameters $\lambda=8.5\times10^{-5}$,
$\chi=2500$, $\varepsilon=280$, $\alpha=0.05$. $Q=100$, $\delta=0$.
Dotted line corresponds to the same values of the parameters but
for ``inverted'' initial conditions
[in Eq.~(\ref{30}) $M_x\rightarrow -M_x$, $M_z\rightarrow -M_z$].
Dashed line represents the dependence $a(\rho)$ with no paramagnetic
moment ($\lambda=0$).}
\label{fig:2}
\end{figure}

To verify our analytical results we solved numerically the
exact equations of motion~(\ref{11}). Fig.~2 (solid line) demonstrates
the dependence of the stationary amplitude of the cantilever vibrations,
$a$, on the frequency detuning, $\rho$. (The stationary amplitude
is achieved at $\tau\gg Q$.) The initial conditions are taken in the form,
\begin{equation}
\label{30}
Z(0)=-1,~ \dot Z(0)=0,~
M_x(0)={\varepsilon\over\sqrt{\varepsilon^2+\chi^2}},
~M_y(0)=0,~
M_z(0)={\chi\over\sqrt{\varepsilon^2+\chi^2}}.
\end{equation}
For these initial conditions at $\tau=0$ the
paramagnetic moment, $\vec M$, points in
the direction of the effective magnetic field, while
the cantilever is displaced in $-z$-direction
($\vartheta_0=3\pi/2$) from its equilibrium position.

\begin{figure}
\centerline{\includegraphics[width=13cm,height=7cm]{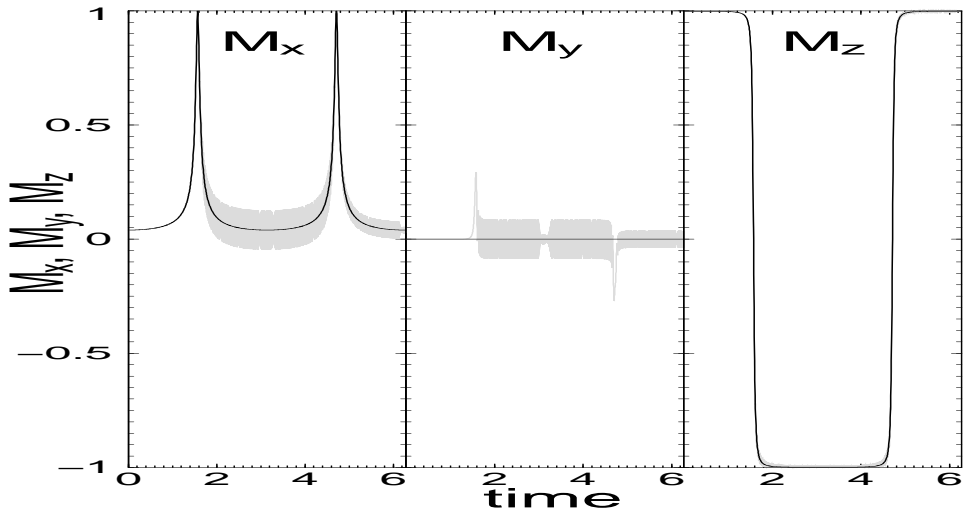}}
\caption{The dynamics of the projections of the paramagnetic moment, $\vec M(\tau)$, of the
sample with the initial conditions (\ref{30}).
The gray line is obtained as a result of the numerical
integration of Eq.~(\ref{11}), and the black line indicates the
quasi-static solution (\ref{13}).
For $M_z(\tau)$ both curves almost coincide. The parameters are the same as those for the solid
line in Fig.~2.}
\label{fig:3}
\end{figure}

Fig.~3 demonstrates the motion of the paramagnetic moment, $\vec M(\tau)$.
One can see the close correspondence between the analytical and numerical
solutions. Note that the frequency shift caused by the adiabatic reversals
changes its sign if the paramagnetic moment points initially
in the direction opposite to the effective magnetic field (while
$Z(0)=-1$). Dotted line in Fig. 2 depicts this case.
We also should note, that decreasing the parameter
$\varepsilon$ (the $x$-component of the effective magnetic field)
leads to the violation of the CAI conditions. Fig. 4 demonstrates
this situation for $\varepsilon=28$.

\begin{figure}
\centerline{\includegraphics[width=13cm,height=7cm]{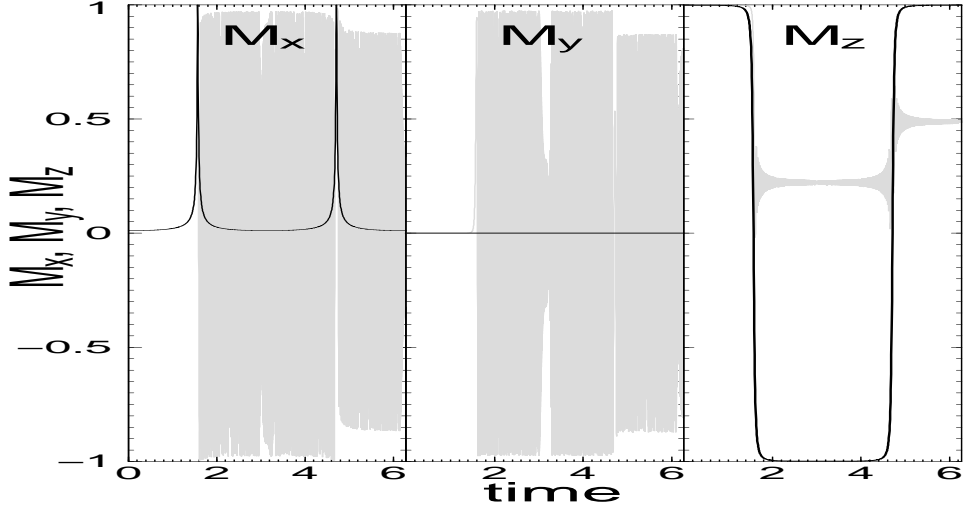}}
\caption{The same as in Fig.~3 but for $\varepsilon=28$.}
\label{fig:4}
\end{figure}

\section{Damped oscillations of the cantilever}
The influence of the sample on the cantilever can be measured
if one turns off the external force acting on the cantilever,
and measures the frequency
of small damped oscillations of the cantilever.
In absence of the paramagnetic moment $\vec M$
the frequency of the oscillations is independent of time
and equals to $\sqrt{1-1/(4Q^2)}$.

We look for the solution of the cantilever vibrations in the form: 
$Z=a(\tau)\cos[\tau+\vartheta(\tau)]$.
Then the dynamical equations for the
slow varying amplitude, $a(\tau)$, and phase,
$\vartheta(\tau)$, in the presence of the sample and in the absence of the external
force take the form,
\begin{equation}
\label{31}
{da\over d\tau}=-{a\over 2Q},
\end{equation}
\begin{equation}
\label{32}
{d\vartheta\over d\tau}=-{1\over 8Q^2}-\frac\lambda{2\pi a}\int_0^{2\pi}
{\chi a\cos^2\psi d\psi\over
\sqrt{\varepsilon^2+(\chi a\cos\psi)^2}}.
\end{equation}
For $p\ll 1$ Eq.~(\ref{32}) can be written as,
\begin{equation}
\label{34}
{d\vartheta\over d\tau}=-{1\over 8Q^2}-
\frac{2\lambda}{\pi a}\left[1-{p^2\over 4}
\left(2\ln\frac 4p-1\right)\right].
\end{equation}
The last term in the right-hand side of Eq.~(\ref{34})
describes a change of the frequency
of small oscillations of the cantilever caused by the adiabatic
reversals of $\vec M$.
From Eq.~(\ref{31}) we have $a(\tau)=a(0)\exp[-\tau/(2Q)]$.
One can see from Eq.~(\ref{34}) that for the initial conditions
(\ref{30}) the influence of $\vec M$
results in decrease of the frequency of small oscillations of the
cantilever in comparison with the case $\lambda=0$.
For small $p$
(the value of $p=\varepsilon/(a\chi)\sim\exp(t/2Q)$ increases with time)
the frequency of oscillations decreases when time increases, as shown
in Fig.~5, while in the absence of the sample this frequency remains
independent of time. We should note that in the studied
approximation the sample does not influence the amplitude of the
cantilever oscillations.

\begin{figure}
\centerline{\includegraphics[width=13cm,height=7cm]{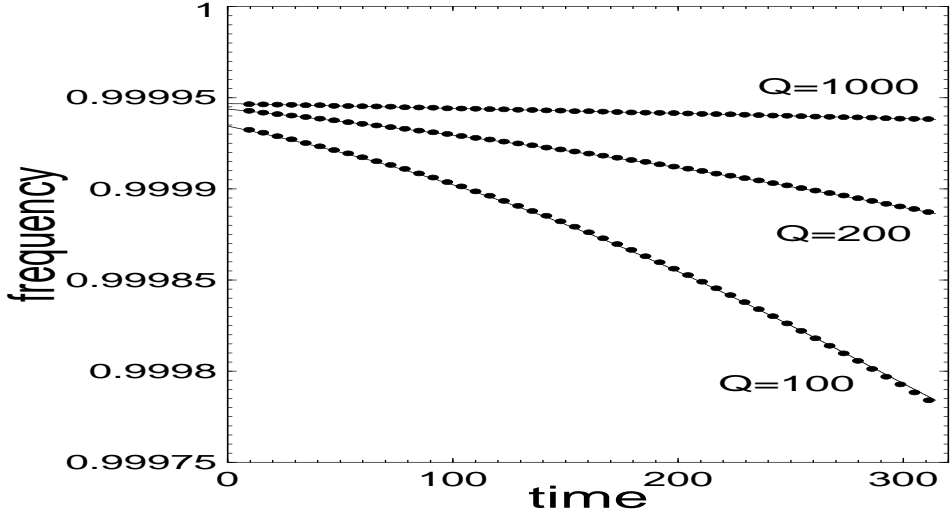}}
\caption{The frequency of small damped oscillations of the cantilever
for the initial conditions (\ref{30}) as
a function of time, $\tau$, for three values of the quality factor, $Q$.
Solid lines are obtained using Eqs.~(\ref{31}) and
(\ref{34}). The results of exact numerical solution are plotted by
the filled circles, $\lambda=8.5\times10^{-5}$, $\chi=2500$,
$\varepsilon=280$, $\delta=0$, $\alpha=0.05$.}
\label{fig:5}
\end{figure}

\section{Summary}
We have studied theoretically and numerically the stationary cantilever vibrations in the
novel OSCAR MRFM technique. Our results are based on the application of
the classical theory for the motion of the cantilever and the paramagnetic
moment of a cluster on the surface of the sample. We have estimated
the resonant frequency shift for the cantilever vibrations.
For the reasonable values of parameters our estimate is about
6 Hz per Bohr magneton. The sign of the shift depends on the
initial direction of the paramagnetic moment relative to
the initial position of the cantilever. We supported our estimation
by the analytical analysis based on the perturbation theory and
by the numerical solution of the equations of motion. Our perturbative
approach is based on the fact that the influence of the paramagnetic
moment on the sample is weak ($\lambda\ll 1$).
We considered also the regime of small oscillations of the
paramagnetic moment near the transversal plane  (linear OSCAR regime).
Finally, we analyzed the damped oscillations of the cantilever
(without the external force). We have shown that the frequency of the damped
oscillations becomes time-dependent due to the adiabatic reversals
of the paramagnetic moment.

\vspace{-5mm}
\section*{Acknowledgments}
\vspace{-3mm}
We are thankful to P.C. Hammel, D.V. Pelekhov, and D. Rugar for useful discussions.
The work was supported by the Department of Energy (DOE) under
contract W-7405-ENG-36, by the National Security Agency (NSA), by the
Advanced Research and Development Activity (ARDA), and
by the Defense Advanced Research Program Agency (DARPA) through the program MOSAIC.

\vspace{-5mm}
\section*{Appendix}
\vspace{-3mm}
Here we express the integral in Eq.~(\ref{21}) in terms of complete
elliptic integrals,
$$
\int_0^{\pi/2}
{\chi a\cos^2\psi d\psi\over
\sqrt{\varepsilon^2+(\chi a\cos\psi)^2}}=
\int_0^{\pi/2}
{\cos^2\psi d\psi\over
\sqrt{p^2+\cos^2\psi}}=
\int_0^{\pi/2}
{(1-\sin^2\psi) d\psi\over
\sqrt{p^2+1-\sin^2\psi}}=
$$
$$
\int_0^{\pi/2}
{(p^2+1)\left(1-{1\over p^2+1}\sin^2\psi\right)-p^2 \over
\sqrt{p^2+1}\sqrt{1-{1\over p^2+1}\sin^2\psi}}d\psi,
$$
where we introduced the notation $p=\varepsilon/(a\chi)$. Splitting this integral in two parts we obtain the right-hand side
of Eq.~(\ref{22}).

{}

\end{document}